\documentclass[10pt]{article}
\usepackage[a4paper,margin=2.25cm]{geometry}
\usepackage{amsmath,amssymb,bm}
\usepackage{graphicx}
\usepackage{booktabs}
\usepackage{siunitx}
\usepackage{microtype}
\usepackage{cite}
\usepackage[hidelinks]{hyperref}
\usepackage{enumitem}
\usepackage{caption}
\title{Physics-constrained identification of near-flutter aeroelastic damping from finite records}
\author{Carlos Domingo M\'endez\\
\small GISPA, Facultad Polit\'ecnica, Universidad Nacional de Asunci\'on, San Lorenzo, Central, Paraguay\\
\small \texttt{cmendez@pol.una.py} \quad ORCID: 0000-0003-3232-9188}
\date{}
\begin{document}
\maketitle

\begin{abstract}
Near flutter, modal frequency can be well resolved from accumulated phase while the sign of a small modal growth rate remains uncertain because the response envelope changes only weakly over a finite record. This work tests whether a validated physical relation between frequency and damping can exploit the better-resolved phase information to improve near-flutter stability identification. A coupling-audited SU2 Euler campaign for a modified Isogai-parameterized NACA 64A010 section at $M_\infty=0.85$ is subjected to systematic grid and temporal verification, multichannel critical-pole tracking, bootstrap trajectory closure, finite-record information analysis, and nonlinear Monte Carlo validation. The temporal study shows that 50 steps per structural period predict the wrong damping sign although the frequency is already close to the finer solutions. The verified critical trajectory gives $V_{\mu,f}=0.537665$ and $q_\omega=d\omega_c/d\sigma_c=3.754$. Restricting a free pole estimate to this trajectory gives an approximately 3.89-fold damping-precision gain, reproduced by Monte Carlo with a median RMSE gain of 3.915. At $|\Delta V_\mu|=10^{-3}$ and 30 dB amplitude SNR, the minimum tested record exceeding 95\% correct-sign probability decreases from 15 to 5 cycles. The results show that physically validated pole trajectories can substantially reduce the observation time required for reliable near-boundary aeroelastic stability identification.
\end{abstract}


\section{Introduction}
Flutter assessment ultimately requires the sign of the real part of a critical aeroelastic eigenvalue. Close to the boundary, the growth or decay rate $\sigma_c$ approaches zero, so a finite record may contain too little envelope evolution to determine whether the mode is stable even when its oscillation frequency is accurately resolved. This difficulty is present in flight-test identification and time-domain computational aeroelasticity, where stability is inferred from subcritical or near-critical modal trends rather than from a single visibly divergent response \cite{Dimitriadis2001,McNamara2007,Mendez2021}. Ground-vibration and pre-flight modal analyses face a related need for robust automated extraction of weakly damped modes \cite{Mendez2025}.

The asymmetry between frequency and damping is physical. Phase accumulates every cycle, whereas near neutrality the exponential factor $\exp(\sigma t)$ changes only slightly. A record can therefore contain tens of cycles and still carry very little direct information about the sign of $\sigma$. This also means that numerical convergence of frequency is not, by itself, evidence of convergence of the stability sign. The temporal verification reported below provides a direct computational example: a coarse time step predicts the wrong damping sign while its frequency is already close to the finer solutions.

Modal estimators based on exponential fitting, matrix pencils, dynamic mode decomposition (DMD), and higher-order DMD (HODMD) provide powerful descriptions of unsteady aeroelastic records \cite{Hua1990,Rowley2009,Schmid2010,LeClainche2017,LeClaincheNoisy2017}. Their flexibility is valuable, but a free near-neutral pole must infer both frequency and damping from the same finite record. Classical Cram\'er--Rao analyses of damped sinusoids quantify this loss of precision \cite{Kay1993,Yao1995,Badeau2008}. The key hypothesis of the present work is that aeroelastic physics can supply useful side information: if a local physical trajectory couples $\omega_c$ and $\sigma_c$, the better-resolved frequency variation can contribute to the determination of the damping sign.

To test this hypothesis without assuming the desired relation, the physical trajectory is first established independently from a verified CFD campaign. The aerodynamic configuration is a modified Isogai-parameterized NACA 64A010 section in the transonic-dip regime, a classical setting for computational flutter studies \cite{Isogai1979,Isogai1981,LeeRausch1995,Reddy1988,Sekar2004}. SU2 \cite{Economon2016} is used for the time-marching Euler solution. The resulting critical-pole trajectory is then used as a constraint in a separate finite-record estimation experiment. The information calculation is used as a quantitative tool rather than as the scientific endpoint; the predicted improvement is independently checked with nonlinear Monte Carlo estimation.

The paper makes four contributions. First, it demonstrates a wrong-sign damping failure caused by insufficient temporal resolution near flutter, despite comparatively small frequency error. Second, it constructs and verifies a local CFD critical-pole trajectory after a solver-level aeroelastic coupling audit and systematic spatial/temporal checks. Third, it quantifies how the physical trajectory reduces finite-record uncertainty and the record length required for stability-sign classification. Fourth, it validates the predicted gain with nonlinear Monte Carlo estimation. A controlled multi-pole information study based on Theodorsen aerodynamics \cite{Theodorsen1935}, together with deliberate trajectory-misspecification tests and optimizer audits, is retained in the Supplementary Information because these analyses support, but do not define, the central aeroelastic result.

\section{Finite-record near-flutter identification}
\subsection{Why frequency can converge before damping sign}
For a local critical response, consider
\begin{equation}
y(t)=e^{\sigma t}\big[a\cos(\omega t)+b\sin(\omega t)\big]+\varepsilon(t),
\label{eq:model}
\end{equation}
where $a$ and $b$ are quadrature amplitudes and $\varepsilon$ represents measurement or effective residual noise. The dimensionless envelope-evolution index
\begin{equation}
\Lambda_\sigma=|\sigma|T_{\rm obs}
\label{eq:lambda}
\end{equation}
measures how much exponential information is present over the observation interval. When $\Lambda_\sigma\ll1$, many cycles may still accumulate phase while the envelope changes by only a few percent or less.

For a near-neutral sinusoid in white Gaussian noise, the leading finite-record scaling is
\begin{equation}
\operatorname{std}(\widehat\sigma)\approx
\frac{\sqrt{24}}{\mathrm{SNR}_A\sqrt{N}\,T_{\rm obs}},
\label{eq:sigscale}
\end{equation}
which implies the one-sided 95\% sign criterion
\begin{equation}
\Lambda_\sigma\gtrsim\frac{8.06}{\mathrm{SNR}_A\sqrt N}.
\label{eq:signcriterion}
\end{equation}
The exact constants depend on the noise definition and covariance structure, but the physical message is general: the available damping information collapses near neutral stability even when the frequency remains measurable \cite{Yao1995,Badeau2008}.

\subsection{Free and physics-constrained pole models}
The free model M0 estimates $(\sigma,\omega)$ independently after profiling $a$ and $b$. The physics-constrained model M1 uses a locally validated aeroelastic trajectory
\begin{equation}
\omega=\omega_f+q_\omega\sigma,
\label{eq:m1}
\end{equation}
where $\omega_f$ is the frequency at the fitted flutter crossing and
\begin{equation}
q_\omega=\frac{d\omega_c}{d\sigma_c}.
\end{equation}
Thus M1 does not infer the stability sign from frequency alone; it uses an independently established physical relation between frequency and damping.

After profiling the quadrature amplitudes, let $\bm F$ denote the effective Fisher matrix for the free pole and $\bm q=(1,q_\omega)^T$ the local trajectory tangent. The free damping variance bound is $C_0=\bm e_\sigma^T\bm F^{-1}\bm e_\sigma$, whereas the locally correct constrained bound is \cite{Stoica1998,Moore2007}
\begin{equation}
C_1=(\bm q^T\bm F\bm q)^{-1}.
\end{equation}
The corresponding standard-deviation gain is
\begin{equation}
G=\sqrt{\left(\bm e_\sigma^T\bm F^{-1}\bm e_\sigma\right)
\left(\bm q^T\bm F\bm q\right)}.
\label{eq:gain}
\end{equation}
For a sufficiently long single-mode record, the profiled metric approaches isotropy in $(\sigma,\omega)$ and Eq.~\eqref{eq:gain} reduces to
\begin{equation}
G\rightarrow\sqrt{1+q_\omega^2}.
\label{eq:isotropic}
\end{equation}
This limit gives the direct physical interpretation used here: a steep validated $\omega$--$\sigma$ trajectory converts the more readily accumulated phase information into damping information. Multi-pole systems can exhibit additional anisotropy; a controlled typical-section example is reported in the Supplementary Information and is not claimed as CFD-validated.

\section{Verified CFD critical-pole trajectory}
\subsection{Modified Isogai-parameterized configuration}
The CFD calculations use SU2~8.5.0 \cite{Economon2016} with the Euler equations for a NACA 64A010 section at $M_\infty=0.85$ and mean angle of attack $\alpha=1^\circ$. The structural parameters follow the widely used Isogai Case-A parameterization: elastic-axis parameter $a_h=-2.0$, center-of-gravity offset $x_\alpha=1.8$, radius of gyration $r_\alpha=1.865$ ($r_\alpha^2\simeq3.48$), mass ratio $\mu=60$, and $\omega_h=\omega_\alpha=100$ rad s$^{-1}$ \cite{Isogai1979,Isogai1981,Sekar2004}. Because the canonical comparison case is commonly evaluated at zero mean incidence, the present configuration is described as \emph{modified Isogai-parameterized}, not as an exact reproduction. Published Euler and Navier--Stokes studies provide the external transonic-flutter context \cite{LeeRausch1995,Reddy1988}; reduced-order CFD approaches provide a complementary route to flutter prediction \cite{SilvaBartels2004}.

The dimensionless flutter speed index is
\begin{equation}
V_\mu=\frac{U_\infty}{b\omega_\alpha\sqrt\mu}.
\end{equation}
Dual-time stepping is second order. The production analysis uses 60 structural periods per operating point and matrix-pencil identification of the critical response in pitch, plunge, $C_L$, and $C_M$.

\subsection{Aeroelastic coupling audit}
A solver-level audit showed that the structural-update frequency and the inner-iteration convergence settings must be coordinated to guarantee an aeroelastic update at every physical time step. Structural motion is updated at the configured aeroelastic inner-iteration cadence; if convergence is permitted before that cadence is reached, a physical time step can terminate without the intended structural update. Production runs therefore retain \texttt{AEROELASTIC\_ITER=5} and impose \texttt{CONV\_STARTITER=20}. Run-level audits confirmed structural updates through every physical step, including the final step. Earlier archived sweeps that did not satisfy this requirement were discarded and are not used quantitatively.

Some post-processing and statistical-analysis scripts were prepared with limited AI-assisted drafting support. All scripts were inspected by the author, executed locally, and independently checked against the frozen numerical datasets and the verification, bootstrap, Monte Carlo, and optimizer tests reported here.

\subsection{Spatial and temporal reliability near flutter}
Because the original hybrid grid did not form a systematic family, a new all-triangular family was generated from the same 128-point airfoil boundary and farfield geometry. The five levels contain 4728, 6580, 9646, 13811, and 21294 points. Their flutter estimates are
\begin{equation}
0.5354354\rightarrow0.5386052\rightarrow0.5434720\rightarrow0.5376909\rightarrow0.5377012.
\end{equation}
The sequence is non-monotonic; therefore no Richardson extrapolation or conventional grid-convergence index is claimed \cite{Celik2008}. The narrower statement is practical stabilization of the finest discretizations: the superfine and ultrafine roots differ by 0.001917\%. The final effective refinement ratio is 1.242 and is reported explicitly. The superfine grid is used for production and the ultrafine grid as an independent fine-grid check.

As an external consistency check, the present ultrafine value $V_{\mu,f}=0.5377012$ is 1.45\% above the approximately $V^*=0.53$ value reported near the canonical $M=0.85$ transonic dip \cite{Sekar2004}. Because incidence and numerical formulations differ, this comparison is not treated as an exact validation.

Temporal verification on the superfine grid at $V_\mu=0.538$ uses 50, 100, and 200 steps per structural period (T50, T100, T200). The pitch growth-rate estimates are $-4.1565\times10^{-3}$, $+3.5091\times10^{-3}$, and $+2.8350\times10^{-3}$ s$^{-1}$, while the corresponding frequencies are 12.94429, 12.97916, and 12.98798 Hz. Thus T50 gives the wrong stability sign although its frequency differs from T200 by only about 0.34\%. Between T100 and T200, $|\Delta\sigma|=6.741\times10^{-4}$ s$^{-1}$ and the relative frequency difference is 0.0679\%; the observed frequency order is $p_t\approx1.98$. T100 is retained for production.

\begin{table}[t]
\centering
\caption{Numerical checks used for the production CFD trajectory.}
\label{tab:vv}
\begin{tabular}{lll}
\toprule
Diagnostic & Result & Interpretation\\
\midrule
Grid sequence & non-monotonic & no Richardson/GCI claim\\
SF--UF flutter difference & 0.001917\% & finest-grid stabilization\\
T100--T200 $|\Delta\sigma|$ & $6.741\times10^{-4}$ s$^{-1}$ & T100 retained\\
T100--T200 $\Delta f/f$ & 0.0679\% & frequency practically stable\\
Observed temporal order & 1.98 & near second order\\
\bottomrule
\end{tabular}
\end{table}

\begin{figure}[t]
\centering
\includegraphics[width=\linewidth]{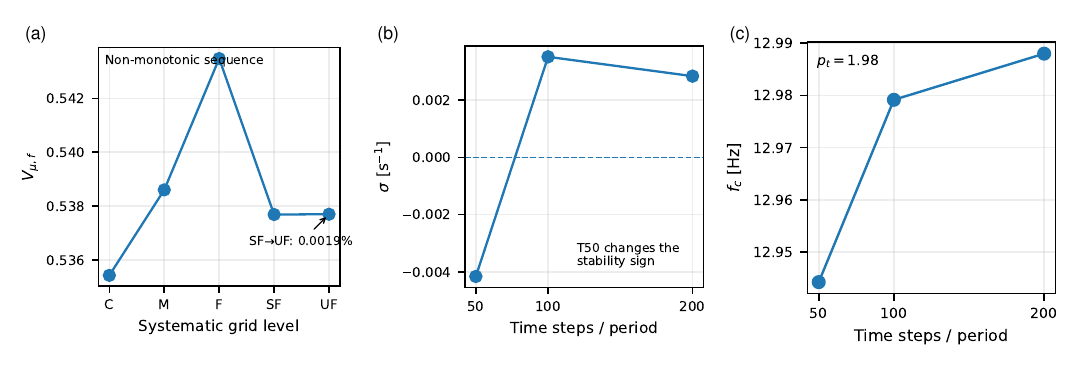}
\caption{Numerical reliability near flutter. (a) The systematic spatial sequence is non-monotonic, followed by close agreement of the two finest discretizations. (b) T50 predicts the wrong damping sign, whereas T100 and T200 agree in sign. (c) The critical frequency is much less sensitive and shows an observed temporal order of approximately 1.98. The central implication is that frequency convergence does not guarantee stability-sign convergence.}
\label{fig:reliability}
\end{figure}

\subsection{Critical-pole trajectory}
The production sweep uses $V_\mu=0.536$, 0.538, 0.540, 0.544, 0.548, and 0.552. Matrix-pencil identification \cite{Hua1990} is applied to multichannel structural and aerodynamic records over a range of ranks. A persistent critical branch is recovered at all speeds. No companion CFD pole is established: an intermittent component near 26 Hz is approximately the second harmonic and is not promoted to an independent aeroelastic mode.

The critical damping and frequency vary almost linearly across the local sweep. The fitted slopes are
\begin{equation}
\frac{d\sigma_c}{dV_\mu}=10.605888\ \mathrm{s^{-1}},\qquad
\frac{d\omega_c}{dV_\mu}=39.815806\ \mathrm{rad\,s^{-1}},
\end{equation}
with $R^2=0.999630$ for damping and 0.999911 for frequency. The crossing is
\begin{equation}
V_{\mu,f}=0.5376647633,
\end{equation}
and the trajectory slope used by M1 is
\begin{equation}
q_\omega=\frac{d\omega_c}{d\sigma_c}=3.7541228.
\label{eq:qomega}
\end{equation}
Leave-one-speed-out fits remain narrow. A paired residual bootstrap that preserves damping--frequency pairing gives a 95\% interval $q_\omega\in[3.67629,3.83082]$ and $\omega_f\in[81.53012,81.54281]$ rad s$^{-1}$. The purpose of these checks is not to manufacture a second branch, but to establish that the observed critical trajectory is sufficiently stable to be used as independent side information.

For the six production records, $\Lambda_\sigma$ from Eq.~\eqref{eq:lambda} is 0.0723, 0.01335, 0.0974, 0.2603, 0.4102, and 0.5710. The $V_\mu=0.538$ record therefore contains 60 oscillation periods but only $\Lambda_\sigma=0.01335$, directly illustrating the weak envelope information near the crossing.

\begin{figure}[t]
\centering
\includegraphics[width=\linewidth]{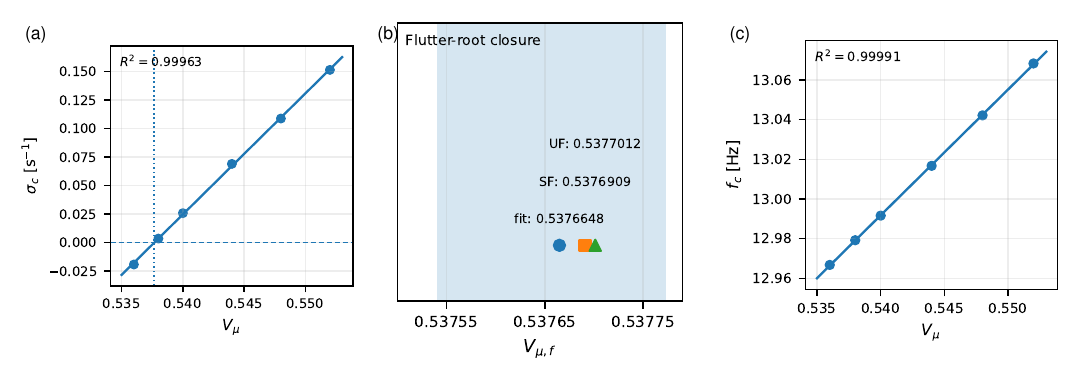}
\caption{Critical CFD pole trajectory. (a) Growth rate versus flutter speed index and the fitted zero crossing. (b) Expanded closure of the trajectory root with the superfine and ultrafine crossings. (c) Critical frequency versus flutter speed index. The trajectory is used only for the observed critical branch.}
\label{fig:trajectory}
\end{figure}

\section{Physics-constrained finite-record stability identification}
\subsection{Physical mechanism and estimation gain}
\begin{figure}[t]
\centering
\includegraphics[width=\linewidth]{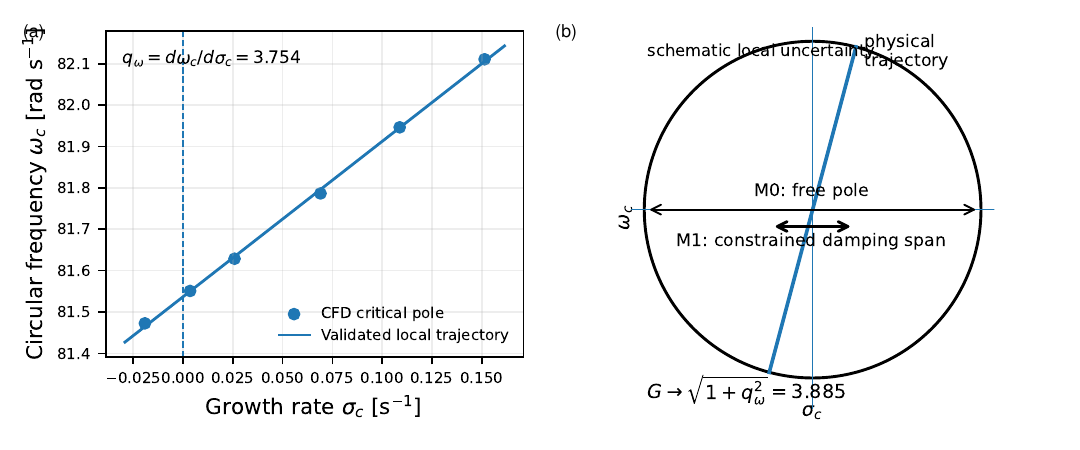}
\caption{Physical-constraint mechanism. (a) The CFD-derived critical trajectory in $(\sigma_c,\omega_c)$ coordinates. (b) Schematic long-record single-mode interpretation: M0 permits a free local pole perturbation, whereas M1 restricts the estimate to the physical trajectory and reduces the admissible damping span. The figure is schematic in panel (b); the numerical slope in panel (a) is obtained from the verified CFD sweep.}
\label{fig:mechanism}
\end{figure}

M0 treats $(\sigma_c,\omega_c)$ as free, whereas M1 imposes the independently obtained local relation
\begin{equation}
\omega_c=\omega_f+3.75412\,\sigma_c.
\end{equation}
Figure~\ref{fig:mechanism} shows both the measured trajectory and the local mechanism. In the long-record single-mode limit, Eq.~\eqref{eq:isotropic} gives
\begin{equation}
G\approx\sqrt{1+3.75412^2}=3.885.
\end{equation}
The interpretation is deliberately simple: the physical trajectory makes a frequency displacement informative about damping because the frequency changes about 3.75 times faster than the growth rate in the local $(\sigma,\omega)$ coordinates. Exact finite-record calculations retain modest phase-dependent anisotropy for the shortest records and rapidly approach this limit.

\subsection{Stability-sign classification}
Correct-sign probabilities are evaluated for $|\Delta V_\mu|=2.5\times10^{-4}$, $5\times10^{-4}$, $10^{-3}$, and $2\times10^{-3}$; amplitude SNRs of 10--40 dB; and records of 2--60 critical-mode cycles. Table~\ref{tab:cycles} reports the minimum tested record length exceeding 95\% correct-sign probability. At $|\Delta V_\mu|=10^{-3}$ and 30 dB, M0 requires 15 tested cycles whereas M1 requires 5. At $5\times10^{-4}$ and 20 dB, the reduction is 60 to 15 cycles. Across the tested grid, the reduction is typically a factor of about two to four.

\begin{table}[t]
\centering
\caption{Minimum tested record length, in critical-mode cycles, required to exceed 95\% correct stability-sign probability. Each entry is M0$\rightarrow$M1.}
\label{tab:cycles}
\begin{tabular}{lrrrr}
\toprule
$|\Delta V_\mu|$ & 10 dB & 20 dB & 30 dB & 40 dB\\
\midrule
0.00025 & $>60\rightarrow60$ & $60\rightarrow30$ & $30\rightarrow15$ & $15\rightarrow5$\\
0.00050 & $>60\rightarrow30$ & $60\rightarrow15$ & $15\rightarrow8$ & $8\rightarrow3$\\
0.00100 & $60\rightarrow30$ & $30\rightarrow15$ & $15\rightarrow5$ & $5\rightarrow2$\\
0.00200 & $30\rightarrow15$ & $15\rightarrow8$ & $8\rightarrow3$ & $3\rightarrow2$\\
\bottomrule
\end{tabular}
\end{table}

A paired bootstrap of trajectory anchor and slope leaves the marginalized discrete threshold grid unchanged, but the closest-to-flutter lower tail is less optimistic. For example, at $|\Delta V_\mu|=2.5\times10^{-4}$, 40 dB, and 5 cycles, the nominal M1 correct-sign probability is 97.77\%, the bootstrap-marginal mean is 95.57\%, and the 2.5th percentile is 87.10\%. Thus the table describes nominal or marginalized local performance, not a worst-case certification guarantee. A deliberately broad trajectory-misspecification analysis is provided in the Supplementary Information.

\begin{figure}[t]
\centering
\includegraphics[width=\linewidth]{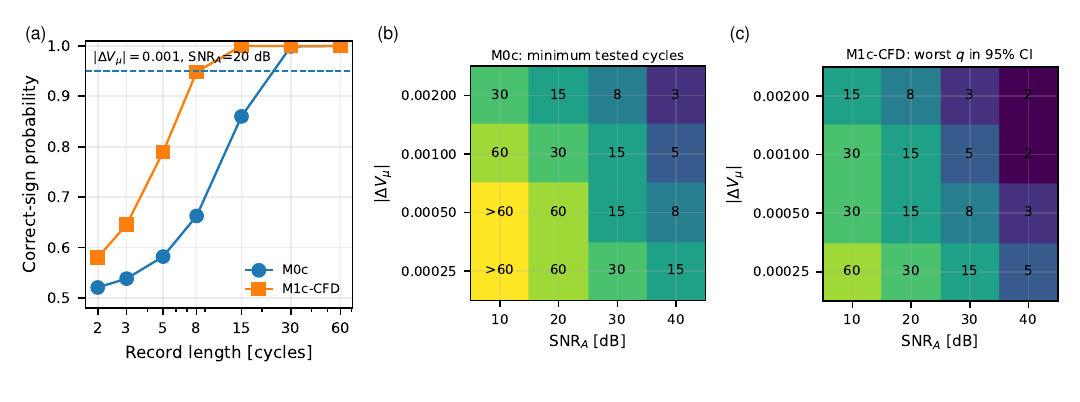}
\caption{Finite-record stability classification. (a) Representative correct-sign probability versus record length. (b,c) Minimum tested cycles required to exceed 95\% for the free and physics-constrained estimators. The physical constraint is beneficial because it couples damping to the better-resolved frequency variation.}
\label{fig:classification}
\end{figure}

\section{Nonlinear Monte Carlo validation}
The information calculation is a local prediction, so it is tested with nonlinear profiled least-squares estimation. Synthetic records use Eq.~\eqref{eq:model} at representative stable and unstable offsets with random phase and additive Gaussian noise. For M0, $(\sigma,\omega)$ are optimized freely while the quadrature amplitudes are solved analytically at each trial pole. For M1, the same profiled likelihood is optimized along the validated trajectory. One thousand realizations per sign and cell are used for the primary validation.

Across the tested cells, the median absolute discrepancy between Monte Carlo and the predicted correct-sign probability is 1.261 percentage points for M0 and 0.259 percentage points for M1. The maximum M1 discrepancy is 0.953 percentage points. More importantly, the empirical ratio
\begin{equation}
\frac{\mathrm{RMSE}_{M0}}{\mathrm{RMSE}_{M1}}
\end{equation}
has a median of 3.915, closely reproducing the predicted 3.885 gain. The stable and unstable sides behave similarly. This agreement shows that the gain is not merely an asymptotic bound artifact over the tested short-record regimes.

\begin{table}[t]
\centering
\caption{Statistical validation of the physics-constrained critical-mode estimator.}
\label{tab:mc}
\begin{tabular}{lc}
\toprule
Quantity & Result\\
\midrule
Validated trajectory slope $q_\omega$ & 3.75412\\
Predicted single-mode gain $G$ & 3.885\\
Median Monte Carlo RMSE gain & 3.915\\
Monte Carlo RMSE-gain range & approximately 3.66--4.08\\
Median $|P_{MC}-P_F|$ for M1 & 0.259 pp\\
Maximum $|P_{MC}-P_F|$ for M1 & 0.953 pp\\
\bottomrule
\end{tabular}
\end{table}

An optimizer audit was also performed on the most difficult cells. L-BFGS-B returned non-success termination flags in 222 of 4000 M0 fits, but all 222 were successfully reoptimized with Powell minimization with zero sign disagreements and no meaningful objective improvement. Changing the M0 damping bounds also changed correct-sign probability by 0.00 percentage points in every audited cell. Details are moved to the Supplementary Information because they verify numerical robustness rather than constitute the primary aeroelastic result.

\begin{figure}[t]
\centering
\includegraphics[width=\linewidth]{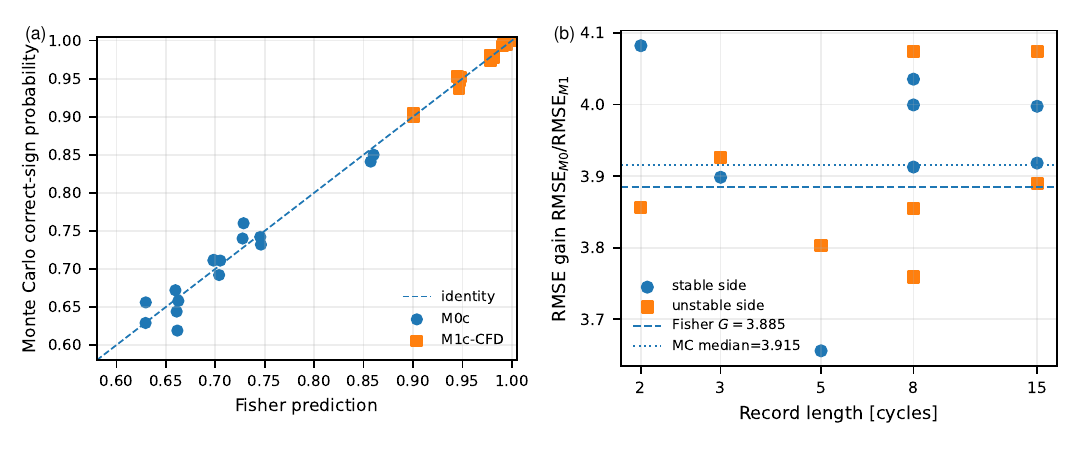}
\caption{Nonlinear Monte Carlo validation. (a) Predicted versus empirical correct-sign probabilities. (b) Empirical RMSE gain for stable and unstable cases compared with the predicted $G=3.885$ line.}
\label{fig:mc}
\end{figure}

\section{Discussion}
\subsection{Frequency convergence is not stability convergence}
The temporal study and the finite-record estimator address two distinct failure mechanisms that lead to the same practical error. Insufficient temporal resolution can shift a near-neutral numerical growth rate across zero even when frequency already appears nearly converged. Separately, an accurately resolved CFD time history can still provide insufficient statistical information to determine the sign of a small physical growth rate. The T50/T100/T200 comparison therefore belongs to the scientific argument, not only to numerical housekeeping: it demonstrates that frequency convergence does not imply stability-sign convergence.

\subsection{Why the physical trajectory helps}
The CFD trajectory gives $q_\omega\simeq3.754$, so a change in growth rate is accompanied by a substantially larger displacement in circular frequency. Because frequency is encoded through accumulated phase, that relation provides side information exactly where the envelope becomes weak. In the observed single-mode CFD case, the long-record information metric is close to isotropic and the gain is therefore governed mainly by the physical slope, $G\simeq\sqrt{1+q_\omega^2}$. This is not presented as a surprising mathematical geometry; it is a quantitative measure of a simple aeroelastic mechanism. In multi-pole systems the information metric can become strongly anisotropic and additional branches can increase the gain, as shown in the controlled Supplementary study. No such extra gain is transferred to the present CFD because a persistent companion pole was not established.

\subsection{Why the trajectory must be independently validated}
A hard constraint can be dangerous if it is wrong. Joint anchor/slope propagation shows that even a well-fitted local trajectory becomes less decisive in the closest-to-flutter lower tail. Deliberate misspecification in the Supplementary Information further shows that a sufficiently wrong slope can bias the estimated damping and eventually reverse its sign. The practical lesson is therefore not that frequency should replace damping, but that a validated physical relation can allow frequency to contribute to damping identification. For experimental use, trajectory uncertainty should be propagated and a soft or probabilistic constraint may be preferable when model-form uncertainty is substantial.

\subsection{Scope and limitations}
The study validates a critical-pole trajectory for one modified Isogai-parameterized transonic configuration. It does not establish a second CFD pole, does not claim formal grid independence, and does not infer a Richardson/GCI uncertainty from the non-monotonic spatial sequence. The superfine--ultrafine agreement and the external $V^*\simeq0.53$ comparison are consistency checks rather than proofs of asymptotic convergence. The analysis also uses Euler aerodynamics; viscous and experimental extensions would test how robust the identified trajectory is to additional physics. These limitations bound the present claim rather than alter it: within the verified local critical branch, physically constrained identification reduces finite-record uncertainty and the effect is reproduced by a nonlinear estimator.

Data-driven flutter prediction and control increasingly use reduced-order or Koopman-based models \cite{Simiriotis2023,Wang2024}. The present result is complementary: before a damping trend is extrapolated or supplied to a controller, the available record must first support its sign. The proposed constraint provides one mechanism for increasing that support when an independent aeroelastic trajectory is available.

\section{Conclusions}
Near a flutter boundary, accurate frequency identification is not equivalent to accurate stability identification. In the verified temporal study, a coarse time step gives the wrong damping sign even though the frequency is already close to the finer solutions, and the near-critical production record at $V_\mu=0.538$ contains 60 cycles but only $\Lambda_\sigma=0.01335$.

A coupling-audited CFD campaign establishes a local critical trajectory with $V_{\mu,f}=0.537665$ and $q_\omega=d\omega_c/d\sigma_c=3.754$. The observed companion CFD pole is not established, so the constraint uses only the critical branch. This separation is essential: the reported gain comes from validated critical-pole information rather than from assumed multi-pole structure.

Restricting the finite-record pole estimate to the physical trajectory gives an approximately 3.885-fold damping-precision gain. In a representative case at $|\Delta V_\mu|=10^{-3}$ and 30 dB amplitude SNR, the minimum tested record exceeding 95\% correct-sign probability falls from 15 to 5 cycles. Nonlinear Monte Carlo estimation independently gives a median RMSE gain of 3.915 and reproduces the constrained sign probabilities to within 0.953 percentage points over the tested regimes.

The practical conclusion is that physically validated modal trajectories can transfer more readily accumulated phase information into damping-sign information and thereby reduce the observation time required for reliable near-flutter classification. The benefit is conditional on the trajectory itself being credible; uncertainty or model misspecification must therefore be propagated rather than hidden.

\section*{Statements and Declarations}
\noindent\textbf{Competing interests.} The author declares no known competing financial interests or personal relationships that could have appeared to influence the work reported in this paper.

\noindent\textbf{Author contribution.} Carlos Domingo M\'endez: Conceptualization, Methodology, Formal analysis, Investigation, Software, Validation, Visualization, Writing -- original draft, Writing -- review and editing.

\noindent\textbf{Data availability.} The frozen numerical-verification summaries, critical-pole trajectory data, finite-record analysis tables, Monte Carlo results, and post-processing scripts can be supplied for peer review. A public archival repository will be prepared for the final accepted dataset.

\noindent\textbf{Acknowledgements.} The author acknowledges the institutional research environment of GISPA and the Facultad Polit\'ecnica, Universidad Nacional de Asunci\'on.

\end{document}